# OUTBREAK: A user-friendly georeferencing online tool for disease surveillance

Raúl Arias-Carrasco[1,#], Jeevan Giddaluru[2,#], Lucas E. Cardozo[2], Felipe Martins[2], Vinicius Maracaja-Coutinho[1,3,*], Helder I. Nakaya[2,4,*]

[1]Advanced Center for Chronic Diseases – ACCDiS, Facultad de Ciencias Químicas y Farmacéuticas, Universidad de Chile, Santiago, Chile
[2]Department of Clinical and Toxicological Analyses, School of Pharmaceutical Sciences, University of São Paulo, São Paulo, Brazil
[3]Instituto Vandique, João Pessoa, Brazil
[4]Scientific Platform Pasteur USP, São Paulo, Brazil
[*]Co-corresponding authors
[#]These authors contribute equally to this work.

**Corresponding author information:**
Helder I Nakaya, PhD
Av. Prof. Lúcio Martins Rodrigues, 370, block C, 4th floor
São Paulo – SP - Brazil
CEP:  05508-020
Phone: + 55 11 2648-1130
e-mail: hnakaya@usp.br

Vinicius Maracaja-Coutinho, PhD
Santos Dumont, 964, Independencia
Santiago – Región Metropolitana – Chile
Código Postal: 8380494
Phone: +56 9 51097362
e-mail: vinicius.maracaja@uchile.cl

**Abstract**

The current COVID-19 pandemic has already claimed more than 100,000 victims and it will cause more deaths in the coming months. Tools that can track the number and locations of cases are critical for surveillance and can help in making policy decisions for controlling the outbreak. The current surveillance web-based dashboards run on proprietary platforms, which are often expensive and require specific computational knowledge. We present a new tool (OUTBREAK) for studying and visualizing epidemiological data. It permits even non-specialist users to input data most conveniently and track outbreaks in real-time. This tool has the potential to guide and help health authorities to intervene and minimize the effects of the outbreaks. It is freely available at http://outbreak.sysbio.tools/.

**Background**

Effective epidemiological surveillance is essential to ensure that the response to infectious disease outbreaks is timely and adequate. Communicable disease surveillance provides the essential information to monitor, evaluate, and model the impact of preventive and control activities; to detect and track the spread of the epidemic and emerging diseases and other threats to public health and to geographically locate communicable diseases in countries, regions, and across the globe. Integrated surveillance enables health authorities to (i) identify populations at risk, (ii) implement prevention and control strategies, (iii) detect unusual disease patterns, and (iv) contain the re-emergence or emergence of communicable diseases [1].

Recent Coronavirus epidemics have displayed the capacity to spread rapidly. The SARS epidemic (2003–2004) infected over 8,000 people in 17 countries with a mortality rate of 9.6% [2]. Later in 2012, MERS-CoV infected 2,519 subjects with a mortality rate of 34.3% in 27 countries [3]. The current SARS-COV-2 pandemic has already claimed more than 100,000 victims and it will cause more deaths in the coming months [4]. Coronaviruses spread through human contact and also through objects contaminated by respiratory droplets exhaled by the infected persons [5]. Because of its rapid spread and high rate of infection [6], by infecting a large proportion of the population it can cause the collapse of the health system of a country [7]. Tools that track and update the number and locations of cases are critical for surveillance of an epidemic and can help the decision-makers to fashion effective policies for controlling the spread of the disease.

Several web-based dashboards display surveillance datasets on the current COVID-19 outbreak, such as by Johns Hopkins University [8], the University of Virginia (http://nssac.bii.virginia.edu/covid-19/dashboard/), HealthMap developed by Harvard University and Boston Children's Hospital (helthmap.org/wuhan), and The World Health Organization's (WHO) dashboard (https://who.maps.arcgis.com/apps/opsdashboard/index.html), among others. Although most of these epidemic dashboards use open-source data, they run on Esir ARCGIS web-services which are often expensive and hard to build and require specific knowledge on GIS-based programs. These dashboards are limited in monitoring the outbreak from a spatial epidemiological perspective, are without a time axis to track the evolution of the epidemic, and they do not allow the user to input data. To overcome these issues, we developed a new web-based tool for studying and visualizing epidemiological data, allowing the user to input data in the most convenient way for non-specialists to track outbreaks in real time and in a user-friendly manner.

## OUTBREAK Overview

OUTBREAK facilitates epidemic surveillance by showing in an animated graph the timeline and geolocations of cases of an outbreak. It is made available in a user-friendly web site (http://outbreak.sysbio.tools/), where even non-specialists can input data in the most convenient way. The tool tracks outbreaks in real-time, and exercises surveillance over the spread of epidemics in regions, countries or, even, across the globe. For example, we built maps and animations showing the evolution of the 2003 SARS epidemic (https://outbreak.sysbio.tools/animation/SARS_2003) and the evolution of the ongoing COVID-19 cases, from January 22 to March 19, 2020 (https://outbreak.sysbio.tools/animation/COVID19).

## Availability and Codes

OUTBREAK online tool is freely available at http://outbreak.sysbio.tools/. The software includes a text and video tutorial with a detailed description of how to use it. OUTBREAK was developed using Python, through the Flask environment [9]. The interface was implemented using JavaScript through the React.js (reactjs.org) with the Node.js library [10]. The map for georeferencing was implemented using the MapBox service through the React.js library "react-map-gl". An up-to-date version of the tool is available for downloading at Docker Hub (https://hub.docker.com/r/integrativebioinformatics/outbreak) together with the information on how to install and run the software locally.

## Data Preparation and Input

OUTBREAK input file requires geographical (latitude and longitude pairs) and temporal (date) information. Users are required to provide this information in a tab-delimited text file. The file can be generated as an Excel sheet that has at least four fixed columns for geographical (Label, Latitude, Longitude) and temporal (Date) data (Table 1). The other three optional columns representing the colors, size, and the number of occurrences for each point can also be included in this file (Table 1). This file can be uploaded at the "Run" page (Figure 1A), after filling a form with the title and a brief description to be shown on the interactive map (Figure 1B). The optional columns in the input file are useful to differentiate datasets of interest to be visualized in the interactive map. For instance, in Figure 1B the reported cases for the COVID-19 outbreak are represented in orange, whereas the associated deaths are represented in red.

## Outbreak Surveillance and Interactive Map Exploration

The tool accepts as input any file containing georeferenced and dated information. Users can specify the dates from a calendar to visualize the outbreak evolution throughout a particular period (Figure 2A). The user can generate animations by hitting the "play" button at the bottom of the animation page (Figures 1B and 2B). It is possible to change the speed of animation and share it on social media. When the animation is displayed, two dynamic graphs are automatically generated, one that shows the number of cases per day and the other that shows the cumulative number of cases till that day. Both graphs evolve according to the period previously selected in the calendar. The graph boxes can be dragged and dropped to facilitate interactive map inspection. Examples of such use of the graphics of SARS 2003 and COVID-19 2020 outbreaks are shown in Figure 2C.

To exercise surveillance over an outbreak, decision-makers and epidemiologists may need to explore the geographical evolution of cases over time and classify the cases by criteria such as incoming cases, ethnicity, sex, and age, among others. To make the datasets graphically clear for exploration, OUTBREAK enables the users to apply different colors for each variable of interest (Figure 3). Finally, the high-level zoom-in feature allows for deeper surveillance of an outbreak in the city of interest at different levels, such as neighborhood, street and even at a single building level (Figure 3). To demonstrate these features, we also displayed the car incidents from San Francisco Open Data Portal (https://datasf.org/opendata/) where each type of incident has a different color (https://outbreak.sysbio.tools/animation/EXAMPLE).

## Conclusion

OUTBREAK is a web-based tool that permits easy surveillance of any epidemiological data. The high level of zoom-in enables the user to track data, at every geographical level—on a worldwide scale, at the level of a city level, a neighborhood, a street or a building. Moreover, the data can be presented graphically in real-time to easily show the daily changes in the number and the spread of cases. The tool only needs world coordinates (in decimal degrees format) and dates to generate the graphics. Besides helping in monitoring the spread of epidemics, OUTBREAK can be used to handle any type of world-wide data, such as for animal migration studies and population studies among others. Finally, we hope that the ease of use and the availability of interactive and dynamic maps that make exploration of the data easy will guide and help health authorities and decision-makers in making effective interventions to minimize the undesirable impacts of the current and future outbreaks.

## Acknowledgments

We would like to thanks Lucas Fleig for his technical assistance in the development of the tool. HIN is supported by CNPq (313662/2017-7) and the São Paulo Research Foundation (FAPESP; grants 2017/50137-3, 2018/14933-2, 2018/21934-5 and 2013/08216-2). VMC and RAC are supported by FONDAP-CONICYT (15130011).

## Tables

**Table 1**–General description of the input file for generating interactive maps in OUTBREAK. A description and necessary titles of mandatory and optional columns are displayed.

| Column title | Data description |
|---|---|
| **Mandatory columns:** | |
| **Label** | ID of respective Latitude and Longitude pairs. |
| **Latitude** | Latitude value (Decimal Degrees) |
| **Longitude** | Longitude value (Decimal Degrees) |
| **Date** | Date in day-month-year format |
| **Optional columns:** | |
| **Color** | The names of user-defined colors for representing data on the interactive map. One color to be assigned to each set of data. |
| **Size** | Size of the points displayed on the interactive map. |

## Figures

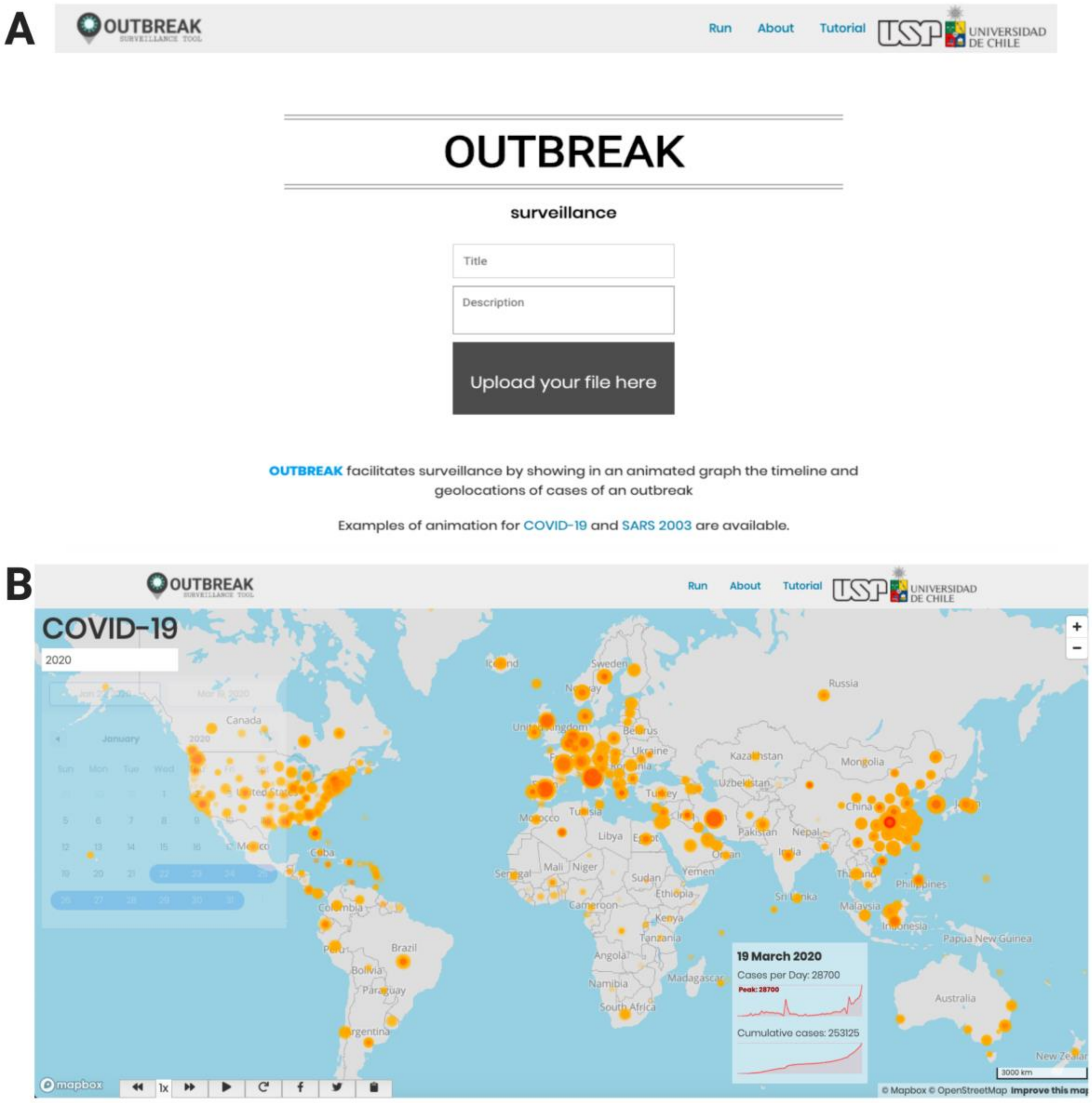

**Figure 1 - General Overview of OUTBREAK Tool. (A)** Run section of the tool, where the user uploads the input file and inserts the title and brief description of the surveillance analysis to be performed. **(B)** Result section, showing the dynamic interactive map generated after running the tool.

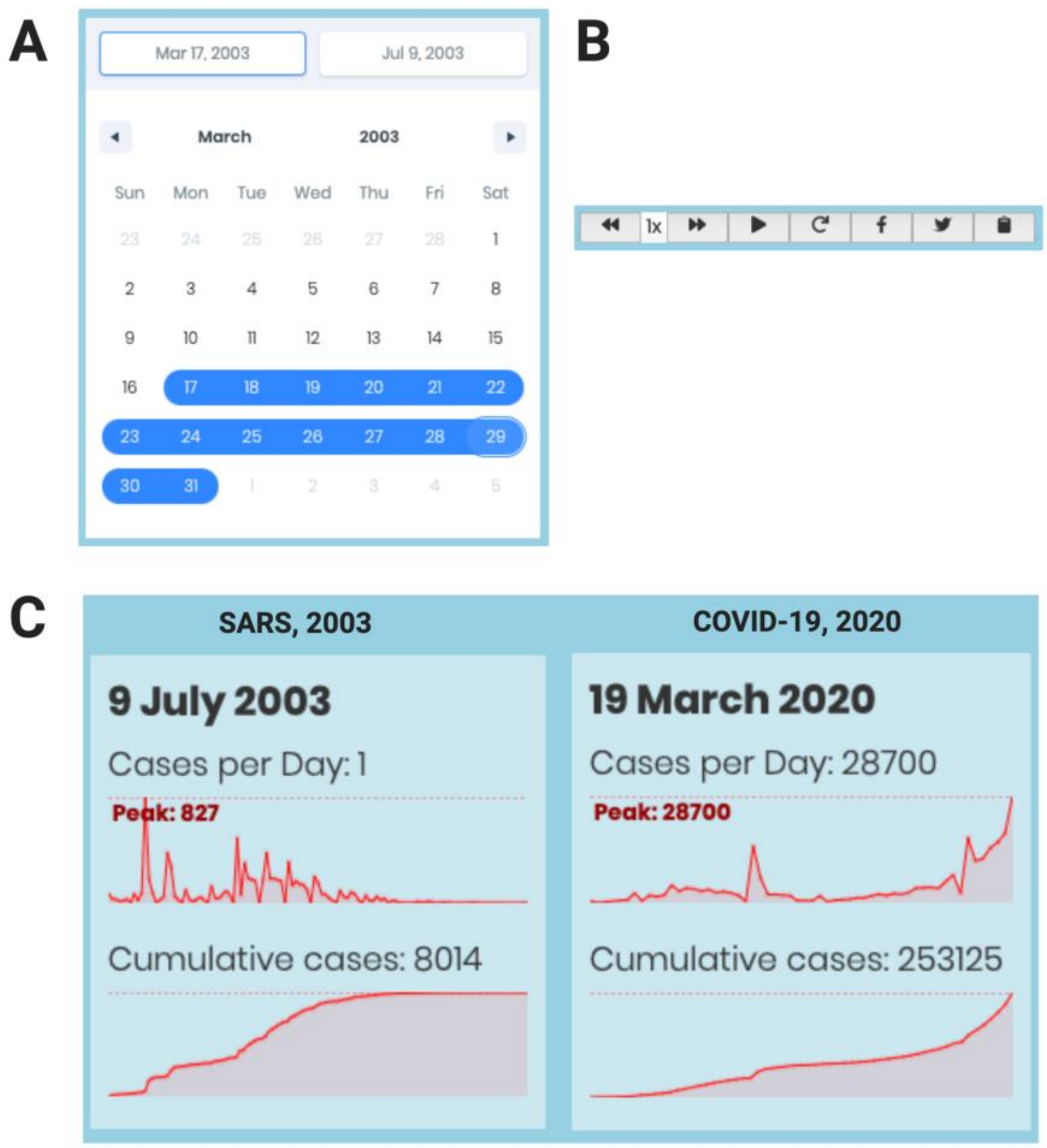

**Figure 2 - Different Features Available in OUTBREAK. (A)** Calendar box used to select a particular period of interest to be explored with the tool. **(B)** Control menu for the dynamic video generated by the OUTBREAK. The user can rewind or fast-forward the animation, with further options to play, pause, restart, or share the video in social media or by copying its URL to the clipboard. **(C)** Dynamic graphics generated when running the tool, showing the cases per day and cumulative cases for SARS 2003 and COVID-19 2020 outbreaks.

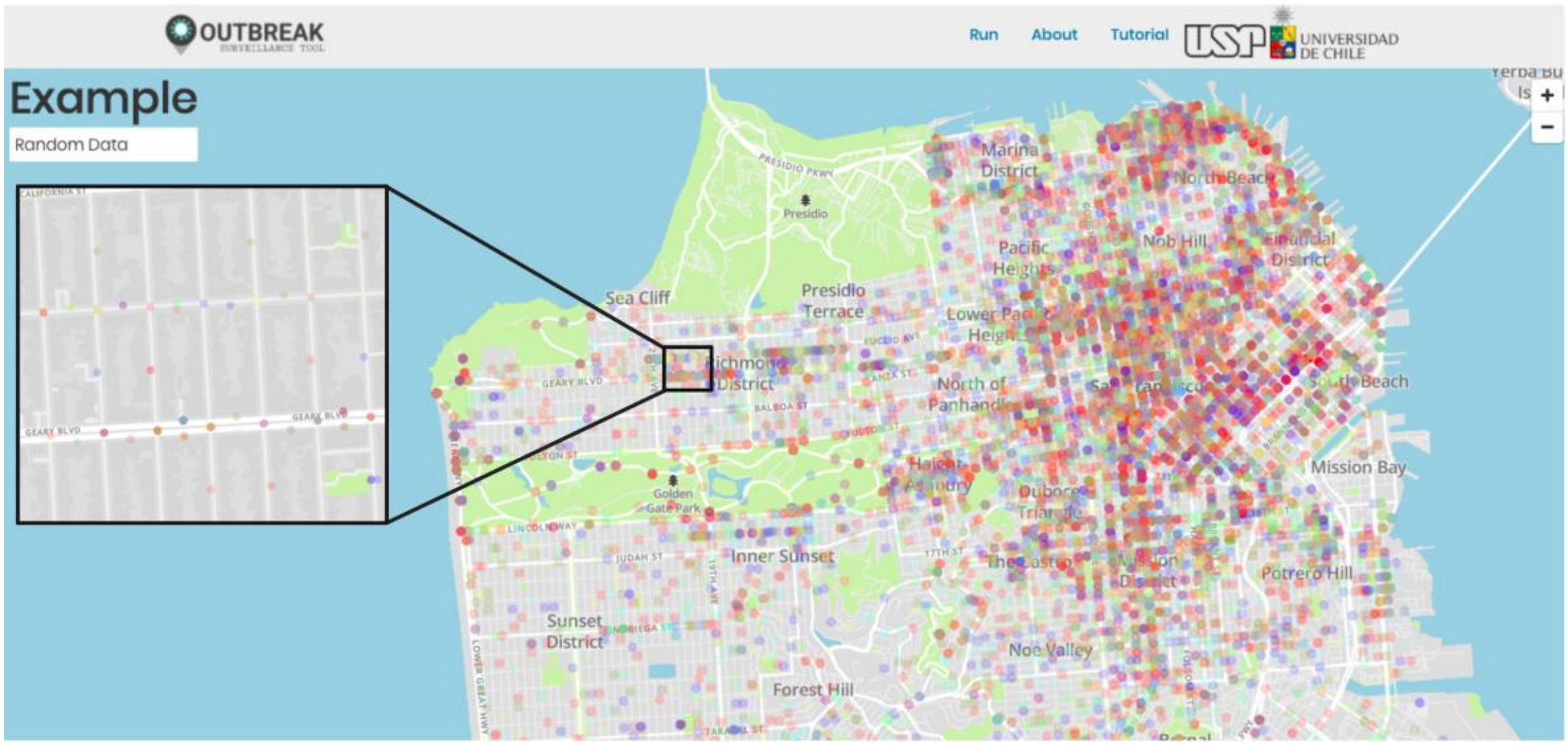

**Figure 3 – Zoom and Color Features.** Example of surveillance of a hypothetical epidemic in San Francisco using OUTBREAK. Some key features of the tool are represented, such as the use of different colors to show the studied cases, and the use of zoom to investigate particular cases and retrieve the information at a neighborhood, street or even at a single building level.